\documentclass[preprint, amsmath,amssymb,nofootinbib]{revtex4}
\usepackage{graphicx}

\newcommand{\myfigure}[2]{\resizebox{#1}{!}{\includegraphics{#2}}}

\begin{document}
\bibliographystyle{unsrt}

\title{Haloes of k-Essence\\ {\small \today}}

\author{C. Armendariz-Picon}
\email{armen@phy.syr.edu}
\affiliation{Physics Department, Syracuse University.}

\author{Eugene A. Lim}
\email{eugene.lim@yale.edu}
\affiliation{Physics Department, Yale University.}

\begin{abstract}
We study gravitationally bound static and spherically symmetric configurations of  k-essence fields. In particular, we investigate whether these configurations can reproduce the properties of  dark matter haloes. The classes of Lagrangians we consider lead to non-isotropic fluids with barotropic and polytropic equations of state.  The latter include microscopic realizations of the often-considered  Chaplygin  gases, which we find can cluster into  dark matter halo-like objects with flat rotation curves, while exhibiting a dark energy-like negative pressure on cosmological scales. We complement our studies with a series of formal general results about the stability and initial value formulation  of non-canonical scalar field theories, and we also discuss a new class of de Sitter solutions  with spacelike field gradients.
\end{abstract}

\maketitle

\section{Introduction}

The recent remarkable progress in cataloging the demographics of our universe is leading us inexorably towards the conclusion that it consists of about 70\% dark energy, 25\% dark matter and 5\% everything else such as radiation and regular baryonic matter \cite{SDSS-WMAP}. The dominant dark energy component, which has a negative pressure, drives the acceleration of the universe \cite{Riess:1998cb,Perlmutter:1998np}; while the pressureless dark matter gravitationally collapses into non-linear objects, forming the backbone of the large-scale structure of the luminous universe we see \cite{WhiteRees:1978}. 

While the standard model of particle physics provides us a with a quite successful description of this 5\% of baryonic matter in the universe, our knowledge of both dark energy and dark matter is tenuous at best.  One of the more promising incarnations of dark energy involves non-canonical scalar fields, also known as  k-fields \cite{armendariz}. They were  first introduced in the context of  inflationary models \cite{k-inflation}, but it was soon realized that they offer great flexibility in designing various cosmological fluids with negative pressure,  generally known as ``k-essence'' \cite{Chiba, Armendariz-Picon:2000dh}. Examples of fluids that can be cast as k-fields include  Chaplygin and Born-infeld gases \cite{Kamenshchik:2001cp,Bento:2002ps}, tachyon matter \cite{tachyon}, phantom energy \cite{Caldwell:1999ew} and ghost condensates \cite{ghost}. 

Conventionally, we think of dark matter and dark energy as two separate entities. However there is a certain elegance in finding a unified model that describes both simultaneously. Although  Chaplygin gases might embody such a unified description \cite{Kamenshchik:2001cp,Bilic:2005sp},  phenomenologically successful models  are hardly distinguishable from  conventional $\Lambda$CDM, as far as the evolution of the background and its linear perturbations are concerned  \cite{SaTeZaWa,BeanDore}. Recently, k-essence fluids have also found explicit application as possible ways to unify dark matter and dark energy in a model proposed by Scherrer \cite{Scherrer:2004au}. The formation of structure in Schrerrer's unified description has been studied by Giannakis and Hu \cite{GiannakisHu}.

Irrespective of the model  under consideration,  dark matter perturbations have to collapse to form  approximately static objects which serve as gravitational potential wells where luminous matter such as stars and galaxies cluster. The standard paradigm to describe such large-scale structure is the halo model \cite{CooraySheth}, which posits that cold dark matter agglomerates into roughly spherical objects called ``haloes''. In this work, we extend the analysis of structure formation in  k-essence models beyond the perturbative regime \cite{caustics,BiLiTuVi}; in particular we are interested in the structure of haloes made of k-essence fluids. Specifically, we study static and spherically symmetric solutions of the Einstein field equations sourced by a number of astrophysically interesting k-essence models. These include the usual barotropic gases considered in standard cosmology ($p\propto \rho$), the polytropic gases ($p\propto \rho^{k}~,~k>0$) conventionally studied in astrophysics and the generalized Chaplygin/Born-Infeld gas models ($p\propto \rho^{-k}~,~k>0$) proposed as forms of dark energy. The final case is especially intriguing, since cosmologically the fluid has negative pressure while in the static regime, as we shall see, it can form spherical objects with roughly flat rotation curves, implying that cosmological perturbations of dark energy may collapse to form dark matter haloes.

Before we proceed, we would like to comment on some philosophical questions about our approach. In general, dark matter is treated as an ensemble of low momentum particles, as opposed to as  a classical field. In other words, given a matter field Lagrangian, we conventionally assume that dark matter is described by an incoherent excitation of  the field quanta,  rather than by a coherent one.   However, in this paper, we take the second approach: we directly solve the Einstein equations  with  k-field matter as a classical source. This is akin to some treatments of bosonic stars \cite{Lynn:1988rb, Jetzer:1991jr} and bosonic dark matter haloes  \cite{Lee:1995af,ArLeSa}. 

This paper is organized as follows. In Section \ref{sec:mainmodel}, we state our model and derive the weak field equations that   spherically symmetric and static configurations of a k-field have to satisfy. In Section \ref{sec:Darkmatter}, we solve these equations for two particular forms of a k-field Lagrangian: a barotropic model (Section \ref{sec:barotropic}) and a polytropic/Chaplygin gas model (Section \ref{sec:polytropeLag}). Finally in Section \ref{sec:Conclusion}, we present our conclusions. In addition, we have included several appendices which contain various derivations of relations and asides, that would have cluttered up the main text. However, we would like to highlight Appendix \ref{sec:stability}, where we provide a careful examination of the conditions one must impose upon the k-field Lagrangian for it to have a valid initial value formulation, and Appendix \ref{sec:nonsingular} where we discuss a new class of de Sitter solutions with spacelike gradients (as opposed to the timelike de Sitter solutions of k-inflation \cite{k-inflation}).

\section{Properties of Inhomogeneous K-essence fluids} \label{sec:mainmodel}

In this section, we introduce the Lagrangian of our model. We write down the field equation of motion and compute its stress-energy tensor, identifying the relevant physical variables along the way. For later use, we comment on the important differences between time-dependent and homogeneous---``cosmological"---  field configurations and its static inhomogeneous counterparts.  As we shall see, while the stress tensor of a cosmological  field does describe a perfect fluid, the stress tensor of a static field does not. Following this discussion, we focus on the equations of motion of static  and  spherically symmetric field configurations by deriving the generalized hydrostatic equations and their weak field limit forms. We postpone the investigation of their solutions to the next section.

\subsection{The Model}
Our starting point is a non-canonical scalar field minimally coupled to gravity \cite{k-inflation},
\begin{equation}\label{eq:action}
	S=\int d^4 x \sqrt{-g} \left[-\frac{R}{16\pi G}+L(X)\right],
\end{equation}
where the Lagrangian $L$ is some arbitrary function of the squared gradient of the scalar field $\varphi$,
\begin{equation}\label{eq:squaredgradient}
	X=\frac{1}{2}\nabla_\mu \varphi \nabla^\mu \varphi.
\end{equation}
Following \cite{armendariz}, we call such a scalar field a ``k-field''. Note that we are working in the $(+,-,-,-)$ metric signature convention. In general $L$ can be a function of both $X$ and $\varphi$, though here we assume that the  Lagrangian only depends on  $X$. This purely $X$-dependent Lagrangian is the most general one compatible with a shift symmetry ${\varphi\to \varphi+const.}$ and second order equations of motion. The shift symmetry implies that there is a conserved current ${dL/dX \cdot\nabla^\mu \varphi}$, whose conservation equation is simply the field equation of motion
\begin{equation}\label{eq:conservation}
	\nabla_\mu\left(\frac{dL}{dX} \nabla^\mu \varphi\right)=0.
\end{equation}
Varying the action with respect to the metric $g_{\mu\nu}$ yields the k-field energy momentum tensor
\begin{equation}\label{eq:EMT}
	T_{\mu\nu}=\frac{dL}{dX}\nabla_\mu \varphi \nabla_\nu \varphi-L(X) g_{\mu\nu}.
\end{equation}
The behavior of such a k-essence fluid depends on the scalar field gradient, which in turn depends on the symmetries of the spacetime the field is living in. In the often considered case of a  k-field in a Friedman-Robertson-Walker (FRW) universe, the symmetry of the Robertson-Walker metric demands that the scalar field be spatially homogeneous (up to small perturbations), so  its gradient is timelike, $X>0$. In this case, the energy momentum tensor has perfect fluid form,
\begin{equation}
T_{\mu\nu}=(\rho_c+p_c)u_{\mu} u_{\nu}-p_c \,g_{\mu\nu},
\end{equation}
where the energy density $\rho_c$ and the pressure $p_c$ are given by
\begin{equation}\label{eq:rho}
	p_c=L,\quad\rho_c=2X \frac{dp_c}{dX}-p_c\quad\quad  (X>0).
\end{equation}
We have added a subscript $c$ for the ``cosmological'' density and pressure terms to distinguish them from their counterparts in the inhomogeneous case below. Note that the four-velocity of the fluid is proportional to the scalar field gradient. In contrast, if the scalar field is static but spatially inhomogeneous, then  its gradient is spacelike, $X<0$, and the energy momentum is \emph{not} of prefect fluid form. Instead, it can be cast as (note the last term)
\begin{equation}
	T_{\mu\nu}=(\rho+p)\,n_\mu n_\nu+\rho\, g_{\mu\nu},
\end{equation}
 where $n^\nu$ is a spacelike unit vector in the direction of the scalar gradient, $n^\mu=\nabla^\mu \varphi/\sqrt{-2X}$.  Hence, an observer whose four-velocity is perpendicular to $n^\mu$ measures an energy density $\rho$, a pressure in the direction perpendicular to $n^\mu$ equal to $-\rho$, and a pressure in the direction parallel to $n^\mu$ equal to $p$, where the energy density and radial pressure are given by
 \begin{equation}\label{eq:p}
 	\rho=-L,\quad p=2X\frac{d\rho}{dX}-\rho \quad\quad (X<0).
 \end{equation}
In other words, the pressure of the fluid is no longer isotropic \cite{armendariz}.  Since dark matter haloes are roughly spherical and static, in this paper we specialize to spherically symmetric and static configurations of these k-fields. For later reference,  we summarize  the classical energy conditions for these field configurations in Table \ref{tab:EC}.  For a definition of the classical energy conditions, the reader is referred to \cite{HawkingEllis, CaHoTr,Onemli:2004mb}.

\begin{table}
\begin{tabular}{|c|c|}
\hline
Weak Energy Condition (WEC) & $\rho\geq 0, \rho+p\geq 0$ \\ \hline
Null Energy Condition (NEC) & $\rho+p\geq 0$ \\ \hline
Dominant Energy Condition (DEC) & $|\rho|\geq |p| \, \text{and WEC}$\\ \hline
Null Dominant Energy Condition (NDEC) & $ |\rho|\geq |p|$ \\ \hline
\end{tabular}
\caption{Energy conditions for a spacelike scalar field gradient.\label{tab:EC}}
\end{table}

\subsection{Spherically Symmetric Static Solutions}
The most general static and spherically symmetric spacetime metric is
 \begin{equation}\label{eq:metric}
 	ds^2=e^{2\alpha(r)}dt^2-e^{2\beta(r)}dr^2-r^2 (d\theta^2+\sin^2\theta d\phi^2),
 \end{equation}
where $\alpha$ and $\beta$ are two $r$-dependent functions. Varying the action (\ref{eq:action}) with respect to the metric leads to the Einstein equation  ${G_{\mu\nu}=8\pi G \, T_{\mu\nu}}$. For the metric (\ref{eq:metric}), $G_{\mu\nu}$ is diagonal, so it follows from the Einstein equation that $T_{\mu\nu}$ also has to be diagonal. Using equation (\ref{eq:EMT}) to compute $T_{tr}$, we find that the first term of the energy momentum tensor (\ref{eq:EMT}), $(dL/dX) \partial_t\varphi \,\partial_r \varphi$, has to vanish. Excluding the special case $dL/dX=0$, we conclude that $\varphi$ only depends either on $t$ or $r$. Solutions where the scalar field only depends on time, $\varphi=\varphi(t)$, have been considered in \cite{BiTuVi}. In this paper, we restrict our attention to solutions where $\varphi$ is only a function of the radius, i.e. 
 \begin{eqnarray}
 	\varphi&=&\varphi(r)\\
        \quad X&=&-\frac{1}{2}e^{-2\beta} \left(\frac{d\varphi}{dr}\right)^2.
 \end{eqnarray}
Then, the non-vanishing terms of the energy momentum tensor are
\begin{subequations}
\begin{eqnarray}
	T^t{}_t&\equiv&\rho=-L \\
	T^r{}_r&\equiv&-p=2\frac{dL}{dX}X-L \\
	T^{\theta}{}_{\theta}&=&T^{\phi}{}_{\phi}=-L,
\end{eqnarray}
\end{subequations}
Note the presence of a significant difference between the radial pressure $p$ and the transverse pressure $L$.  The next step in our derivation is to write down the relations of hydrostatic equilibrium for the system, which are the various components of the Einstein equation. However, for the applications we have in mind, it is not necessary to consider the full non-linear Einstein equations. Indeed, for most astrophysical situations the weak field limit conditions
\begin{subequations} \label{eq:Weakfield}
 	\begin{eqnarray}
 	\frac{G m}{r}&\ll& 1, \\
 	4\pi G\, p\, r^2 &\ll& 1
	\end{eqnarray}
\end{subequations}
apply. Then, if we define
 \begin{equation}\label{eq:beta}
 	e^{-2\beta}=1-\frac{2G m}{r}, \quad \text{or} \quad \beta\approx \frac{G m}{r},
  \end{equation}
we find that the Einstein equations yield the following set of structure equations
\begin{subequations}\label{eq:TOV}
\begin{eqnarray}
	 \frac{dm}{dr}&=&4\pi \,\rho\, r^2, \label{eq:dm} \\
	\frac{d\alpha}{dr}&=&\frac{G m+4\pi G\, p\, r^3}{r(r-2 G m)}
	\approx \frac{G m+4\pi G p\, r^3}{r^2},\label{eq:dalpha} \\
	\frac{dp}{dr}&=&-\left(\frac{2}{r}+\frac{d\alpha}{dr}\right)(\rho+p)
	\approx -\frac{2}{r}(\rho+p),\label{eq:motion}
 \end{eqnarray}
 \end{subequations}
where the equal sign denotes the exact relations, while the approximation denotes their weak field regime cousins. The first two equations are the $tt$ and $rr$ components of the Einstein equation respectively, while the 3rd equation is a combination of the $rr$ and the $\theta\theta$ components. These ``master" equations are very similar to the Tolman-Oppenheimer-Volkov equations of stellar structure \cite{Wald}, where matter is assumed to be a perfect fluid. Since the k-essence fluid is not perfect, there is an additional $2/r$ term in equation (\ref{eq:motion}) arising from the tangential pressure terms. Note that (\ref{eq:motion}) is equivalent to the equation of motion (\ref{eq:conservation}). Thus, this set of equations suffices to determine all the unknowns in the problem, which are $\alpha$, $\beta$ and $\rho=-L$ ($p$ is related to $L$ in terms of equation (\ref{eq:p})). 

\section{k-essence Fluids as Dark Matter Haloes} \label{sec:Darkmatter} 

Since the Lagrangian $L(X)$ is an arbitrary function, the space of possible field configurations which satisfy the Einstein equations  is infinite. The only statement we can make about the fluid  at this moment is that it is ``isentropic'', by which we mean that  the pressure is simply a function of its density $p=p(\rho)$. We are, however, interested in functions that can lead to dark matter halo-like configurations. To this end, we focus on two particular forms of the Lagrangian which are both simple enough to solve and have such properties. As we discuss next, our choices of Lagrangians can be also motivated as follows.
 
The macroscopic behavior of a fluid is described by a set of state parameters. This set of parameters determines its stress energy tensor, which is then coupled to gravity via Einstein's equations. These parameters are in general not independent, but constrained instead by an equation of state, which depends on the  microscopic properties of the fluid. In many cases, the equation of state can be written in the following form
\begin{equation}\label{eq:spolytropic}
	p=K\rho^{\gamma}
\end{equation}
where $K$ and $\gamma$ are constants. If $\gamma=1$, we call such a fluid \emph{barotropic}, and if $\gamma\neq 1$ we call it \emph{polytropic}. An example of an isotropic barotopic fluid is a bath of photons where $K=1/3$, while an example of a polytropic fluid is a bath of non-relativistic degenerate electrons of a white dwarf star, where $\gamma=5/3$. Both the above examples, and other isotropic fluids, have been extensively studied in the field of stellar structure and evolution \cite{Chandrasekhar}. In this paper, we investigate the structure of \emph{non-isotropic} k-essence fluid haloes. In analogy with equation (\ref{eq:spolytropic}) we postulate the equation of state for the k-essence fluid  to be of the form (\ref{eq:spolytropic}) where $p$ and $\rho$ are the pressure and energy density defined in equation (\ref{eq:p}). Then, it follows that for homogeneous configurations of the scalar field, the equation of state in a FRW universe (if defined) has the inverse form,
\begin{equation}
	\left[p\propto\rho^{\gamma}\right]_{\mathrm{static}} \leftrightarrow \left[p_c\propto \rho_c^{1/	\gamma}\right]_{\mathrm{homogeneous}}.
\end{equation}
One can see this simply by comparing the relations between $p$ ($p_c$) and $\rho$ ($\rho_c$) in the two cases, namely equations (\ref{eq:rho}) and (\ref{eq:p}), and noting that the roles of $p$ ($p_c$) and $\rho$ ($\rho_c$) are interchanged between the two. Furthermore, we show in Appendix \ref{sec:stability} that this inversion also applies to the squared sound speed of its perturbations $c_s^2\equiv \delta p/\delta\rho$, as one would naively expect.

Given the form of the equation of state (\ref{eq:spolytropic}), we can work backwards and derive the k-field Lagrangian we need to describe such a fluid. It turns out that the corresponding Lagrangians span simple polynomials or non-analytic functions already proposed in quite different contexts.  We show in Appendix \ref{sec:nonsingular} that, regardless of the Lagrangian, the only solution to the equations of motion regular at the origin is de Sitter space. Therefore, the solutions we discuss next are not well-defined for all values of $r$, and they generically become singular as $r\to 0$. Since we expect new physics as this limit is approached, we shall not be concerned about this fact. In any case, because  we only study the weak field limit of the gravitational equations, we do not encounter this singular behavior in the regime of validity of our approximations. 

The solutions we find this way are characterized by two boundary conditions, the energy density $\rho_0$ and the mass $m_0$ at a given, arbitrary value of the radius $r_0$.   Presumably, their value is determined by the environment that lead to the formation of the halo, a process we cannot address  with our static solutions. Since we cannot determine the value of these parameters, and because they determine some of the halo observables, like their mass, we mainly focus on the scaling of our solutions and on their main properties.

\subsection{Barotropic Lagrangians}\label{sec:barotropic}

We first consider a barotropic k-essence fluid $p\propto \rho$.  Suppose that the Lagrangian $L$ is an analytic function of $X$. Then, $L$ can be expanded in a power series,
\begin{equation} 
L=M^4 \sum_n c_n \left(\frac{X}{M^4}\right)^n,
\end{equation}
where the $c_n$ are dimensionless coefficients and $M$ is an arbitrary mass scale.  For a given value of $X$, we expect one particular term of the series to dominate, depending on the values of $c_n$. Hence, we can assume that (at least for a certain range of values of $X$) the Lagrangian is given by the monomial
\begin{equation}
	L=M^4 \left(\frac{X}{M^4}\right)^n\label{eq:baroLag}.
\end{equation}
If the scalar field is spatially homogeneous, as it must be if it shares the symmetries of  a FRW spacetime, such a scalar field behaves as a perfect fluid with a barotropic equation of state 
\begin{equation}
p_c=w\rho_c,
\end{equation}
where for convenience we have defined the equation of state parameter
\begin{equation}
w=\frac{1}{2n-1}.
\end{equation}
In contrast, as we have described in Section \ref{sec:mainmodel}, for static and spherically symmetric configurations the inverse relation holds
\begin{equation}\label{eq:barotropic}
	p=\frac{\rho}{w}.
\end{equation}
Similarly, inserting the Lagrangian (\ref{eq:baroLag}) into the sound speed equation (\ref{eq:cssq}), we find that in the static case the sound speed is given by
\begin{equation}\label{eq:barocssq}
	c_s^2=\frac{1}{w},
\end{equation}
as expected from the equation of state (\ref{eq:barotropic}). The speed of sound is superluminal if $w<1$ ($n>1$.) As detailed in Appendix \ref{sec:stability}, this is not a violation of causality, since the k-field  propagates on the ``light-cone'' of a different metric.   In fact, unless $w=1$, the speed is superluminal for either homogeneous or static field configurations. Equation (\ref{eq:barocssq}) also shows that if  $w<0$ ($n<1/2$), the k-field is unstable to perturbations. Notice that in order for the energy density to be positive in both cosmological and static cases $n$ has to be odd.

In order to compute the energy density profile of a halo, we insert equation (\ref{eq:barotropic}) into the weak field structure equations (\ref{eq:TOV}), and  find that the energy density is a simple power law in $r$,
\begin{eqnarray}\label{eq:barsolution}
	\rho&=&\rho_0 \left(\frac{r}{r_0}\right)^{-2(1+w)} \\
	m&=&m_0+\frac{4\pi r_0^3 \rho_0}{1-2w}\left[\left(\frac{r}{r_0}\right)^{1-2w}-1\right],
\end{eqnarray}
where $\rho_0$ and $m_0$ are the energy density and mass at $r_0$. The energy density $\rho$ is positive if $\rho_0$ is positive. Now we divide the solutions into three different classes: ${n<1/2}$, ${1/2<n\leq 3/2}$  and ${n>3/2}$. We will discuss each in that order, with the last case being the most astrophysically interesting.

In the first case, $n<1/2$,  the equation of state parameter $w$ is negative, leading to a negative squared sound speed $c_s^2<0$ via equation (\ref{eq:barocssq}). Thus, configurations with $w<0$ are unstable, and we shall not consider them here except to mention that they do not violate the weak field conditions as $r\to 0$.

Before we proceed to the next range, we note as an aside that $w=\infty$ ($n=1/2$) is a special case; using equation (\ref{eq:barotropic}) we see that the radial pressure $p$ identically vanishes for this $n$. Although this is not compatible with the weak field equation (\ref{eq:motion}), it is a valid solution of the full equations of motion if $d\alpha/dr=-2/r$.  In that case the equations of motion are exactly solvable, and we find that $r$ is a timelike variable. Hence, this solution does not describe a static configuration. 

The second case, $1/2<n \leq 3/2$, implies that $1/2 \leq w <\infty$, which is the only range of $w$ where the mass of the halo remains finite as $r\to \infty$. Note that in order for the weak field approximation to remain valid as $r\to \infty$, $w$ has to be positive  ($n\geq 1/2$), which means that all finite mass solutions can be made consistent with the weak field approximation by an appropriate choice of initial conditions. On the other hand, if we try to extrapolate our solutions to small radii, we find that the week field limit is violated as $r\to 0$. 

Finally, we discuss the case where $n>3/2$, implying that $w<1/2$. This class of solutions, similar to the $1/2 < n \leq 3/2$ case above, violates the weak field limit as $r\to 0$. The mass of their haloes also diverges\footnote{This does not mean that they are unphysical however. As we mentioned above, if the  Lagrangian is a polynomial in $X$, rather than a monomial, different monomials dominate at different values of $X$. At large values of $X$, high powers are relevant, and for low values of $X$, low powers are dominant. Now, because the energy density monotonically decreases as the radius $r$ increases, far from the halo center we expect the Lagrangian to be dominated by a low-power monomial (typically the linear term, $n=1$.) These low-power terms then render the mass of the halo finite. In other words, as $r\to \infty$, one can expect the solution to return to that of the second case.} as $r\to\infty$.  As described above, the equation of state of our barotropic fluid in a cosmological space time is $p_c = w\rho_c$. Thus if $n$ is very large, $w$ is close to zero, which is simply the equation of state for cosmological dark matter. Furthermore, the cosmological sound speed for this matter is $c_s^2=w$ which also approaches $0$ as $n$ tends to large values. Hence, in the linear regime this Lagrangian exhibits all the properties of plain vanilla dark matter. In a dark-matter dominated universe, linear perturbations grow with the scale factor. Once they reach the non-linear regime they collapse into dark matter haloes, which we identify here with the static spherically symmetric matter configurations we have found. As we have shown in equation (\ref{eq:barsolution}), for large $n$ their energy density profile at large $r$  becomes
\begin{equation}
	\rho\approx\rho_0\left(\frac{r}{r_0}\right)^{-2}
\end{equation}
which is just that for a singular isothermal sphere. The singular isothermal sphere is a good approximate model for a galactic halo, because its rotation curves\footnote{i.e. the circular velocity of objects, say stars, orbiting around the galactic center.} flatten out as the radius tends to infinity, as consistently observed in galaxy rotation curves \cite{McGaugh:2001yc}. Since the observed luminous mass is insufficient to explain such a velocity profile, this fact is usually attributed to the existence of non-luminous (cold) dark matter.

Rotation curve observations are usually measurements of the Doppler shift of the light emitted by HII regions or, for nearby galaxies, stars. As shown in Appendix \ref{sec:Rotcurve}, the apparent rotation velocity of an object at radius $r$ inside the k-essence halo, $v(r)$, is given by  equation (\ref{eq:rotcurve}),
\begin{equation}
	v^2=\frac{Gm(r)}{r}+4\pi G r^2 p(r). \label{eq:doppler2}
\end{equation}
For the barotropic model, the right hand side of equation (\ref{eq:doppler2}) is analytically solvable. By plugging in the density profile equation (\ref{eq:barsolution}) into the equation above, we find
\begin{equation}
	v^2(r)\approx 4\pi G r_0^2\rho_0 \frac{1-w}{w-2w^2}\left(\frac{r}{r_0}\right)^{-2w},
\end{equation}
which as $w\to 0$  ($n\to\infty$) becomes a constant, i.e. the rotation curve becomes flat at large $r$. While this is an interesting model of dark matter, as it exhibits the dark matter-like properties cosmologically and in the non-linear regime, it is nevertheless not a very natural model, since $n$ has to be extremely large. Indeed, if the sound of speed dark matter is non-vanishing,  density perturbations are prevented from growing by the non-vanishing pressure, which causes density perturbations to oscillate instead.  The absence of these oscillations in the (galaxy) power spectrum  down to  scales of around 1 Mpc \cite{2dF}  hence  restricts the speed of sound of dark matter to be less than about $10^{-3}$, which translates into $n>10^6$.

The properties of barotropic solutions are summarized in Table \ref{tab:barotropic}. Note that the solutions go into the strong field regime either as $r \to 0$ or $r \to \infty$.  
\begin{center}
\begin{table}
\begin{tabular}{|c|c|c|c|}
\hline
{} & $n<1/2$ & $1/2<n\leq 3/2$ & $ n> 3/2$ \\ \hline
{} & $w<0$   & $1/2<w<\infty$   & $0<w<1/2$ \\  \hline
mass & - & finite & diverges  \\ \hline
valid weak field regime  & $ r\to 0$ & $r\to \infty$ & $r\to \infty $ \\ \hline
$c_s^2$ &$ <0$ &  $>0$  &$ >0$   \\ \hline
\end{tabular}
\caption{Properties of barotropic solutions.\label{tab:barotropic}}
\end{table}
\end{center}

\subsection{Polytropic Lagrangians} \label{sec:polytropeLag}

In this subsection we consider a class of non-analytic Lagrangians that lead to a polytropic equation of state
\begin{equation}
	p=p_* \left(\frac{\rho}{\rho_*}\right)^\gamma, \label{eq:poly}
\end{equation}
where $\gamma$, $p_*$ and $\rho_*>0$ are constants. By inserting equation (\ref{eq:poly}) into equation (\ref{eq:p}), one finds that such k-essence fluids are given by a Lagrangian of the form \cite{Bilic:2001cg,Bento:2002ps,Armendariz-Picon:2003ht,Diez-Tejedor:2005fz}
\begin{equation}\label{eq:polL}
	L=-\left[(-M^4X)^\frac{1-\gamma}{2}-\frac{p_*}{\rho_*^\gamma}\right]^{\frac{1}{1-\gamma}}.
\end{equation}
The first minus sign (in front of the bracket) guarantees that the halo energy density is positive, while the second one (inside the parenthesis) guarantees that the Lagrangian is well-defined for spacelike $X$ and arbitrary $\gamma$.  From this choice of signs it also follows that $L_{,X}$ is positive whenever the expression inside the brackets is positive. As shown in Appendix \ref{sec:stability}, this  is one of the conditions for the quantum stability of the configuration. If in addition, $c_s^2>0$, the configuration is both classically and quantum-mechanically stable.  In order to compute the speed of sound, we insert  the Lagrangian (\ref{eq:polL}) into the sound speed equation (\ref{eq:cssq}) and  find that the sound speed in the static case is
\begin{equation}
	c_s^2=\gamma\frac{p}{\rho} \quad\quad (X<0). \label{eq:soundspeedpoly}
\end{equation}
Since by assumption and construction $\rho>0$, this means that the sign of $\gamma p$ must be positive in order for the perturbations to be stable. This result allows us to rule out many of the possible cases in our following discussion of polytropic solutions. 

The class of Lagrangians (\ref{eq:polL}) is interesting for several different reasons. For instance, if $\gamma=-1$ and $p_*<0$, it becomes the scalar Born-Infeld Lagrangian \cite{BornInfeld}. The Born-Infeld Lagrangian has found many applications in high energy physics and cosmology \cite{Gibbons}. It naturally appears in the description of moving branes \cite{Bilic:2002vm}, and in particular, it has been argued to describe the dynamics of open string tachyons \cite{Sen}. The latter have been also suggested to behave as a pressureless fluid, i.e. as dark matter \cite{tachyon}. If the scalar field is homogeneous, the Lagrangian (\ref{eq:polL}) leads to the polytropic equation of state
\begin{equation}\label{eq:homstate}
	p_c = (-\rho_*)\left(\frac{\rho_c}{-p_*}\right)^{1/\gamma}<0.
\end{equation}
Insisting that the energy density be positive leads to a negative cosmological pressure, a key requirement for an accelerating universe. Note that the speed of sound in this case is
\begin{equation}
	c_s^2=\frac{1}{\gamma}\frac{p_c}{\rho} \quad \quad (X>0),
\end{equation}
which is dark matter-like (small and positive) if $\gamma\ll -1$, even when $p_c\approx-\rho_a$.

For negative $\gamma$ (which includes the Born-Infeld case), the equation of state of a homogeneous k-field  (\ref{eq:homstate}) agrees with the one of  a generalized Chaplygin gas. Chaplygin gases have been widely considered as a unified description of dark matter and dark energy \cite{Kamenshchik:2001cp}. An advantage of our approach is that the Lagrangian (\ref{eq:polL}) provides a microscopic description of a Chaplygin gas. In particular, it shows that in some cases is not sufficient to specify the equation of state of a homogeneous component; one must specify the full Lagrangian to obtain a description of all its possible configurations. In our realization, the Chaplygin ``gas" does not behave as a perfect fluid when  the k-field is static.

Inserting the equation of state (\ref{eq:poly}) into the equation of motion (\ref{eq:motion}) we find 
\begin{equation}\label{eq:polysol}
	\rho=\rho_0\left[\left(1+\frac{\rho_0}{p_0}\right)\left(\frac{r}{r_0}\right)^{\frac{2-2\gamma}{\gamma}}-	\frac{\rho_0}{p_0}\right]^{\frac{1}{\gamma-1}},
\end{equation}
where $p_0$ and $\rho_0$ are the pressure and density respectively at $r_0$. Note that the signs of $p_0$ and $\rho_0$ must follow that of $p_*$ and $\rho_*$ respectively. The expression inside the parenthesis of equation (\ref{eq:polysol}) vanishes at the radius
\begin{equation}
	r_S=r_0\cdot \left(\frac{\rho_0}{\rho_0+p_0}\right)^{\frac{\gamma}{2-2\gamma}}. \label{eq:surface}
\end{equation}
The behavior of the density profile depends on the the position of this point, which depends on the values of $p_0$ and $\gamma$; say, if $\rho(r_S)$ vanishes, then $r_S$ is a ``boundary'' of the halo. Since $p_0$ fixes the overall sign of the pressure, equation (\ref{eq:soundspeedpoly}), stability  requires that $\gamma p_0>0$. On the other hand, the sign of $p_0$ is determined by $p_*$, so that the Lagrangian itself (as opposed to the initial conditions) is what fixes the properties of the solutions. In the following we discuss the different relevant cases, $\{p_*>0,\gamma<0\}$ and $\{p_*<0,\gamma>0\}$, separately. The results are summarized in Table \ref{tab:polytropic}.

\begin{figure}[ptbh]
\myfigure{4in}{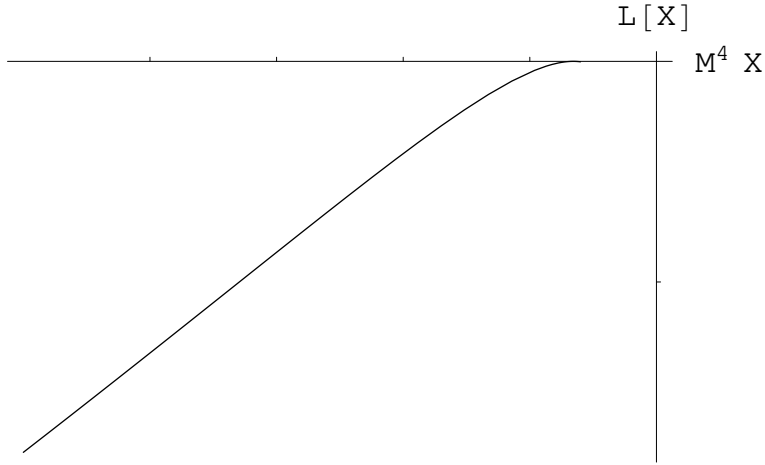}
\caption{A plot of the  Lagrangian (\ref{eq:polL})  for  $p_*>0$ and $0<\gamma<1$ (case \emph{a}). The zero occurs at $X=-M^{-4}(p*/\rho_*^{\gamma})^{2/(1-\gamma)}$, and the pole of $\rho(r_S)$ corresponds to the $X\to -\infty$ limit. Note that the Lagrangian is not defined for positive values of $X$. }
\label{figL1}
\end{figure}

\begin{figure}[ptbh]
\myfigure{4in}{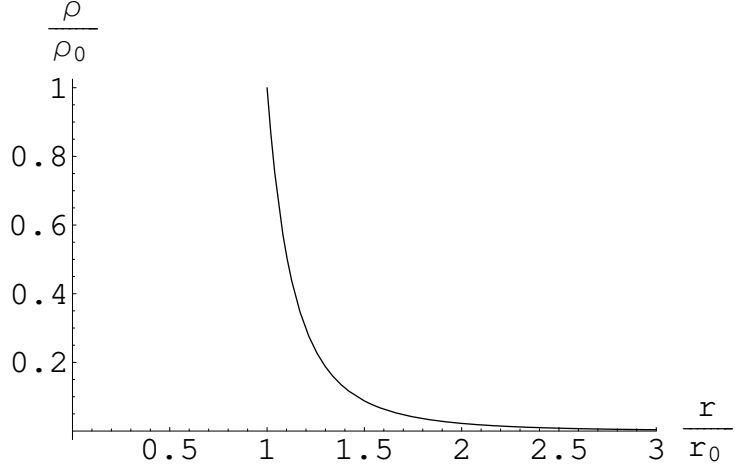}
\caption{The density profile $\rho$ for a  model with $p_*>0$ and $0<\gamma<1$ (case \emph{a}). The parameters are  $p_*/\rho_*=0.9$ and $\gamma=1/2$. Since $\gamma<2/3$, this halo has a finite mass.}
\label{figrho1}
\end{figure}

\paragraph{$p_*>0$ and $0<\gamma<1$}
The energy density has a pole at $r_S<r_0$, and as $r\to\infty$ it  decays as $\rho\propto r^{-2/\gamma}$. Hence, the total mass of the halo is finite for $\gamma<2/3$. Note that the ratio $p/\rho\rightarrow \infty$ as $r\rightarrow 0$, violating the energy conditions listed in Table \ref{tab:EC}. However, these solutions are both classically and quantum-mechanically stable, so they are perfectly sensible scalar field configurations. The Lagrangian for this case is plotted in Fig. \ref{figL1} while its density profile is plotted in Fig. \ref{figrho1}.

Again, we test such models against  observations of the motion of objects around a galactic halo by computing the rotation curves of test particles. At large $r$ the density of such a model becomes
\begin{equation}
	\rho(r)\approx \rho_0\left(1+\frac{\rho_0}{p_0}\right)^{1/(\gamma-1)}\left(\frac{r}{r_0}\right)^{-2/\gamma}.
	\label{eq:largerdensity}
\end{equation}
Thus, in order to calculate the behaviour of rotation curves at large $r$, we  simply insert equation (\ref{eq:largerdensity}) into equation (\ref{eq:rotcurve}),
\begin{equation}\label{eq:rotcurve2}
	v^2(r)=\frac{G m(r)}{r}+4\pi G r^2 p(r).
\end{equation}
Focusing on the first term, we see that the mass of the halo grows at most linearly in $r$, so this term is constant or decays with $r$. Meanwhile the second term of equation (\ref{eq:rotcurve2}) is
\begin{equation}
	4\pi G r^2 p = 4 \pi G r_0^2 p_0\left(1+\frac{\rho_0}{p_0}\right)^{\gamma/(\gamma-1)}
\end{equation}
which is a constant. Hence, the rotation curve $v(r)$ approaches 
\begin{equation}
	v^2(r)= \frac{Gm(r)}{r} + 4\pi G r_0^2 p_0\left(1+\frac{\rho_0}{p_0}\right)^{\gamma/(\gamma-1)},
\end{equation}
which is constant for sufficiently big $r$. As mentioned above, the flatness of stellar rotation curves is conventionally assumed to be due to the presence of dark matter which, by definition, is pressureless. In our halo, the contribution of the  energy density to the rotation curve vanishes or becomes a constant at large $r$, and it is the non-vanishing radial pressure what keeps the rotation curve flat.

\begin{figure}[ptbh]
\myfigure{4in}{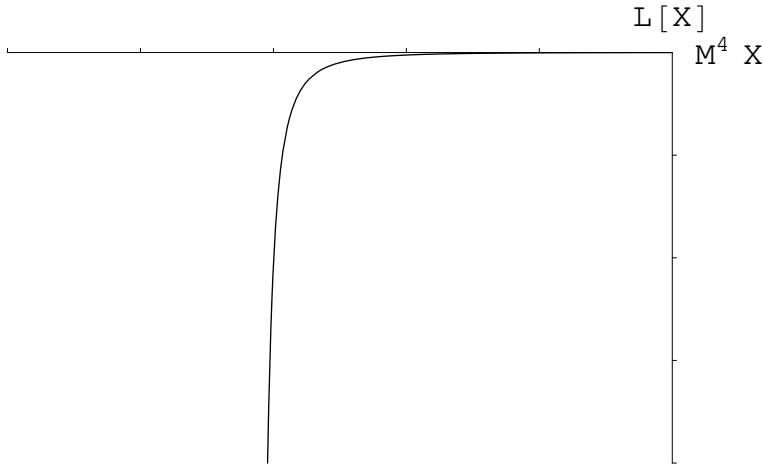}
\caption{Figure showing the Lagrangian (\ref{eq:polL}) for $p_*>0$ and $\gamma>1$ (case \emph{b}). The Lagrangian has a pole at $X=-M^{-4}(p_*/\rho_*^{\gamma})^{2/(1-\gamma)}$, and a zero at $X=0$. Note that the Lagrangian is defined for positive values of $X$ only if $\gamma$ is an odd integer, with the resulting plot being a mirror image of $X<0$.}
\label{figL2}
\end{figure}

\begin{figure}[ptbh]
\myfigure{4in}{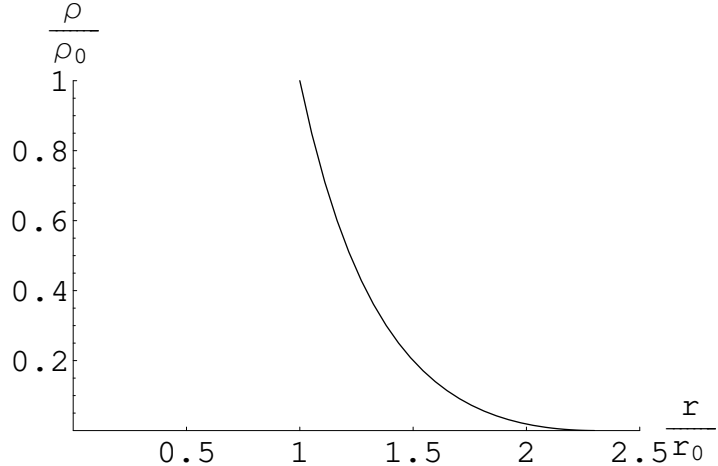}
\caption{Figure showing the density profile for a model with $p_*>0$ and $\gamma>1$ (case \emph{b}). The sample parameters are $p_*/\rho_*=3/4$ and $\gamma=3/2$. There is a point of zero density at finite $r$, and the solution can be joined to the vacuum at that point. }
\label{figrho2}
\end{figure}

\paragraph{$p_*>0$ and $\gamma>1$.}
The Lagrangian and density profile corresponding to this case are shown in Figs \ref{figL2} and \ref{figrho2} respectively. For these values of the parameters, the energy density vanishes at $r_S$. At this point of vanishing energy density, the solution can be joined to the vacuum, $\rho_S=p_S=0$. Therefore, this  halo has a surface, and it seems more appropriate to call it a ``k-star''. Spherically symmetric configurations of (canonical) classical scalar fields have been widely considered in the literature under the name of ``boson stars" (see e.g. \cite{Jetzer:1991jr}.) The difference between these solutions and ours are that in the former the energy density smoothly approaches zero at infinity, while in the latter the energy density vanishes at a finite radius. We shall not analyze the properties of these stars here, since the only relevant observable, their mass, depends on the initial conditions, especially the core mass. However,  these configurations could still play a secondary role in the  dark matter puzzle as massive compact halo objects (MACHOS) \cite{Gerhard:1995ff}. 


\begin{figure}[ptbh]
\myfigure{4in}{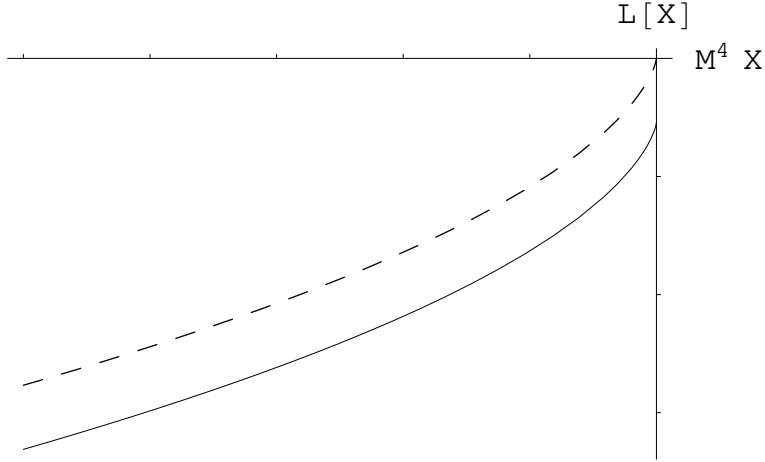}
\caption{Figure showing the Lagrangian (\ref{eq:polL}) for $p_*<0$ and $\gamma<0$ (case \emph{c}). We plot both the original Lagrangian (full line) and the renormalized Lagrangian (dashed line). The asymptotic value of $\rho(r)$ corresponds to the energy density at  $X=0$, which is a non-zero constant (full line). Note that the Lagrangian is defined for positive values of $X$ only if $\gamma$ is a negative odd number. If $(1-\gamma)/2$ is even, the plot of the Lagrangian for $X>0$ is a mirror image of the one for $X<0$. On the other hand, if $(\gamma-1)/2$ is odd, the Lagrangian is differentiable at $X=0$ and vanishes at a finite positive value of $X$.}
\label{figL3}
\end{figure}

\begin{figure}[ptbh]
\myfigure{4in}{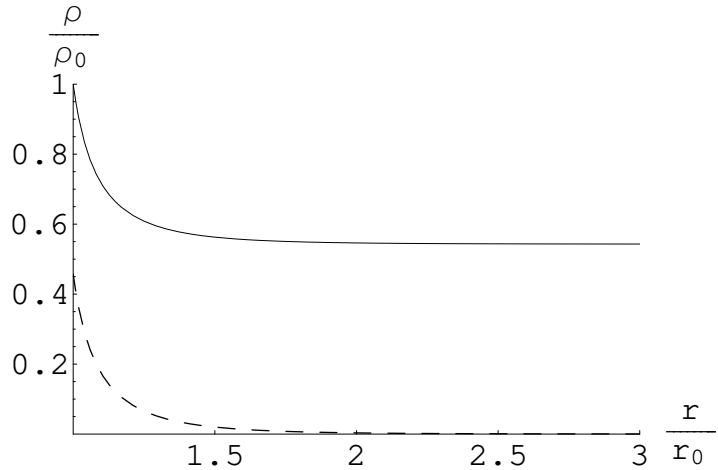}
\caption{Figure showing the density profile for a model with $p_*<0$ and $\gamma<0$ (case \emph{c}). The sample parameters are  $p_*/\rho_*=-0.4$ and $\gamma=-1/2$. The density profile asymptopes at large $r$ to a constant (full line), which can be subtracted off by adding a cosmological constant to the Lagrangian (dashed line), as we do in the text. Note that the $r/r_0$ axis begins at unit value.}
\label{figrho3}
\end{figure}

\paragraph{$p_*<0$ and $\gamma<0$.}
Finally, we consider the case of a generalized Chaplygin gas, which has been suggested as a unified description of  dark matter and dark energy. Since the Lagrangian (\ref{eq:polL}) implies that $p_0/\rho_0>-1$, the energy density has a pole at $r_S<r_0$, and  as $r\to \infty$, it approaches  the constant (see Fig. \ref{figrho3})
\begin{equation}
	\Lambda=\left(-\frac{\rho_*^\gamma}{p_*}\right)^\frac{1}{\gamma-1}.
\end{equation}
Therefore, in this form solutions asymptotically approach de Sitter spacetime (one can also check that the pressure approaches $-\Lambda$), which is what we expect, since Chaplygin gases lead to cosmic acceleration.  From the point of view of the Lagrangian, which we plot in Fig. \ref{figL3}, this is easy to understand. As shown in the Figure, at $X=0$ the Lagrangian is negative and linear in $X$, and hence, it is not surprising to find solutions that asymptotically approach de Sitter. In order to disentangle the vacuum energy from the halo density, we substract this cosmological constant from the Lagrangian,   $L\rightarrow \tilde{L}+\Lambda$ , where  $\tilde{L}$ is our original polytropic Lagrangian. Then, the total density and pressure become $\rho = \tilde{\rho}-\Lambda$ and $p = \tilde{p}+\Lambda$ respectively, where tilded quantities are associated with the original Lagrangian $\tilde {L}$. Because this shift leaves the weak field equation of motion (\ref{eq:motion}) invariant, the new density profile $\rho$ is simply the original profile $\tilde{\rho}$ shifted by the same constant,
\begin{equation}
	\rho=\rho_0\left[\left(1+\frac{\rho_0}{p_0}\right)\left(\frac{r}{r_0}\right)^
	{\frac{2	-2\gamma}{\gamma}}-\frac{\rho_0}{p_0}\right]^{\frac{1}{\gamma-1}}
	-\Lambda.
\end{equation}
Thus one can see that by adding the constant $\Lambda$ to the Lagrangian, we have renormalized the asymptotic value of the energy density at infinity to zero.


In order to compute rotation curves for this halo, we expand the root around its asymptotic value, obtaining 
\begin{equation}
	\rho=\rho_0\left[\frac{1}{1-\gamma}\left(1+
	\frac{p_0}{\rho_0}\right)\left(\frac{\rho_0}{-p_0}\right)^
	{\frac{1}{\gamma-1}}\left(\frac{r}{r_0}\right)^{\frac{2-2\gamma}{\gamma}}+...\right]>0.
\end{equation}
Similarly, the pressure term is approximately given by 
\begin{equation}
p=p_0\left[\frac{\gamma}{1-\gamma}\left(1+\frac{p_0}{\rho_0}\right)\left(\frac{\rho_0}{-p_0}\right)^{\frac{\gamma}{\gamma-1}}\left(\frac{r}{r_0}\right)^{\frac{2-2\gamma}{\gamma}}+...\right]>0.
\end{equation}
Note that the total pressure is positive, though $\tilde{p}$ is negative. Recall now that the rotation curve is given by equation (\ref{eq:rotcurve})
\begin{equation}
	v^2(r)=\frac{Gm(r)}{r}+4\pi G r^2 p(r).
\end{equation}
The pressure  scales like $r^2 p\sim r^{2/\gamma}$, and the mass is $m(r)\sim m_0+m_1 r^{1+2/\gamma}$, where $m_0$ and $m_1$ are integration constants. Therefore, the rotation curve has the form
\begin{equation}
	v^2(r)=\frac{G m_0}{r}+Nr ^{2/\gamma},
\end{equation}
where $N$ is a constant too. Hence, if $\gamma>-2$, the rotation curve goes as $1/r$, and if $\gamma<-2$ the rotation curve decays as $r^{2/\gamma}$ at large r. 

It follows from our considerations that polytropes with $\gamma\ll-1$ are particularly interesting. On one hand,  their rotation curves are flat, so these models correctly reproduce the structure of dark matter haloes.  On the other hand,  their cosmological sound speed is also dark matter-like,  and it turns out that unified models based on these Chaplygin gases  successfully reproduce the linear structures we observe in the universe \cite{SaTeZaWa}.

\begin{center}
\begin{table}[h!] 
\begin{tabular}{| l ||c|c|c||c||}
\hline
	{}	 
	& \multicolumn{2}{|c|}{$p_*>0$}
 	& $p_*<0$ \\ \hline
   		{} &  $0<\gamma<1$ & $\gamma>1$ & $\gamma<0$ \\ \hline
		$r_S$ & $<r_0$ & $>r_0$ & $>r_0$ \\ \hline 
		 $\rho(r_S)$ & pole & zero & pole \\ \hline
		 $r\to\infty$ & $-$ & $\rho\to r^{-2/\gamma}$ & constant \\ \hline
\end{tabular}
\caption{Polytropic solutions. }\label{tab:polytropic}
\end{table}
\end{center}

\section{Conclusions} \label{sec:Conclusion}

We have investigated static and spherically symmetric  solutions to the  Einstein field equations coupled to k-essence.  Our results shed light on the formation of non-linear structures in models where matter can be microscopically described by a non-canonical scalar.  In contrast to homogeneous k-essence fluids, the energy momentum for these configurations does not have perfect fluid form. Rather, it describes an anisotropic fluid with different pressures in the radial and tangential directions. Therefore, the propagation speed of perturbations around these solutions differs from their cosmological counterpart. We have computed the sound speed of the former and  formulated the necessary conditions for these solutions to be both classically and quantum-mechanically stable. 

By making simple assumptions about the form of the k-essence equation of state, we have found solutions to several models of k-fields which exhibit dark matter halo-like properties, namely, flat rotation curves. The corresponding Lagrangians are simple (as in the barotropic case), and physically motivated (as in the Chaplygin gas/polytropic case). In both cases, rotation curves are flat  when the cosmological speed of sound of k-essence tends to zero. The latter include proposed unifications of dark matter and dark energy based on Born-Infeld and generalized Chaplygin  gases. Among these, the ones that  yield flat rotation curves are precisely the ones consistent with the way linear structures forms in our universe \cite{SaTeZaWa}. 

Our solutions are valid only in the limit where the gravitational field of the halo is weak; as one approaches the center of the halo, the weak field limit is generically violated as the density diverges. In other words, haloes made of this form of k-essence fluid also contain a central ``cusp'', whose existence in real haloes is still under debate (see e.g. \cite{cusps}.) In any case, this is a generic problem in common cold dark matter models, and is conceivably resolved only when we add baryonic matter to our model.

Dark matter is a specter in search of an identity. Although simple and appealing, the assumption that dark matter is made out of self-interacting particles requires involved numerical simulations to extract predictions about non-linear structure formation.  In this paper we have explored  an alternative---whether dark matter could be described by a non-canonical scalar field. As we have shown, this explanation has its own advantages. It makes simple, direct  and straightforward predictions that can be immediately compared with observations.

\begin{acknowledgments}
We would like to thank Sean Carroll, Richard Easther, Stefan Hollands, Andrey Kravtsov, Wayne Hu, Mark Trodden and Bob Wald for useful and insightful discussions. E.A.L. is supported in part by DOE-FC02-92ER-40704.
\end{acknowledgments}

\appendix

\section{Rotation Curves} \label{sec:Rotcurve}

Rotation curves of dark matter haloes are determined by measuring the Doppler shift of light emitted by the orbiting objects, such as stars or gaseous (``HII'') regions. In order to compare our haloes with these observations, we study how the light emitted by these objects at radius $r$ is red or blue-shifted. In this appendix, our goal is to derive the rotation curves of massive particles in circular orbits around  k-essence haloes (see also \cite{Nucamendi:2000jw}). 

Consider the motion of massive test particle, say a star, in a such a halo. Its trajectory is then described by a curve $x^{\mu}=(t,r,\theta,\phi)$ parameterized by some affine parameter; here we use its proper time $\tau$. Its four velocity is then simply $u^\mu\equiv dx^\mu/d\tau$. Due to spherical symmetry, we can assume without loss of generality that the star's ecliptic is located in the $\theta=\pi/2$ plane. Since the star is a massive particle, its norm is $u_\mu u^\mu=1$, which becomes the constraint equation
\begin{equation}\label{eq:normalization}
	e^{2\alpha}\dot{t}^2-e^{2\beta}\dot{r}^2-r^2\dot{\phi}^2=1,
\end{equation}
where a dot denotes a derivative with respect to proper time $\tau$. Since the metric does not explicitly depend on $\theta$, the star's angular momentum $J$ is conserved,
\begin{equation}\label{eq:J}
	J=r^2\dot{\phi}.
\end{equation} 
Similarly, the metric does not explicitly depend on $t$, and there is a conserved energy $E$,
\begin{equation}\label{eq:E}
	E=e^{2\alpha}\dot{t}.
\end{equation}
Substituting equations (\ref{eq:E}) and (\ref{eq:J}) into equation (\ref{eq:normalization}), we find a first integral for the motion of the star,
\begin{equation}
	\frac{1}{2}\dot{r}^2+V(r)=0, \label{eq:dop1}
\end{equation}
where its effective potential is
\begin{equation}  
	V(r)=\frac{1}{2}e^{-2\beta}\left(1+\frac{J^2}{r^2}\right)-\frac{1}{2}E^2 e^{-2(\alpha+\beta)}. 
		\label{eq:dop2}
\end{equation}
Note that the potential explicitly depends on the energy. Stationary orbits at radius $r_e$ exist if $V$ and $dV/dr$ vanish at that radius.  The former condition yields 
\begin{equation}
	1+\frac{J^2}{r_e^2}=E^2 e^{-2\alpha(r_e)},
	\label{eq:dopcond1}
\end{equation}
whereas  the latter gives us
\begin{equation}
	-\frac{d\beta}{d r}\left(1+\frac{J^2}{r_e^2}\right)
	-\frac{J^2}{r_e^3}+E^2\left(\frac{d\alpha}{d r}+\frac{d \beta}{d r}\right)e^{-2\alpha(r_e)}=0
	\label{eq:dopcond2}.
\end{equation}
Substituting equation (\ref{eq:dopcond1}) into equation (\ref{eq:dopcond2}) and using the 2nd TOV equation (\ref{eq:TOV}), we get the following equation 
\begin{equation}
	\frac{J^2/r_e^2}{1+J^2/r_e^2}=\frac{Gm(r_e)}{r_e}+4\pi G\, r_e^2\, p(r_e),
	 \label{eq:doppler}
\end{equation}
which directly relates the angular momentum $J_e$ to the density profile of the halo. 

With the angular momentum in hand, we turn to the task of finding the red or blue-shifting of light emitted by an object a such an orbit. Consider a photon with momentum $p^{\mu}_e$ emitted by an orbiting star at radius $r_e$, with its three-momentum assumed to be parallel to the star's three-velocity, 
\begin{equation}\label{eq:photonp}
	p^\mu_e=\left(p^t_e,0,0,e^{\alpha(r_e)} \frac{p^t_e}{r_e}\right).
\end{equation}
Since the star can move towards us or away from us, its four-velocity is given by
\begin{equation}\label{eq:obsu}
	u^\mu=\left(E e^{-2\alpha(r_e)},0,0,\pm \frac{J}{r_e}\right).
\end{equation}
Meanwhile, the frequency of the emitted light $\nu_e$ is given by $\nu_e=p^\mu_e u_\mu$, and thus, by using equations (\ref{eq:photonp}) and (\ref{eq:obsu}) we find the zero component of the photon momentum is
\begin{equation}
	p^t_e=\nu_e\cdot \left(E\mp \frac{e^{\alpha(r_e)} J}{r_e}\right)^{-1}.
\end{equation}

Let us now turn to the computation of the frequency measured by a stationary observer far away from the halo, at $r_\infty$. Since she is stationary, her four-velocity is $v^\mu=(e^{-\alpha(r_\infty)},0,0,0)$, so this observer measures the photon  frequency to be $\nu_\infty=p^\mu_\infty v_\mu=e^{\alpha(r_\infty)}p^t_\infty$.  To determine $p^t_\infty$, we use the conservation of $p_t=e^{2\alpha}p^t$, which yields the red/blue-shifted frequency the observer would measure
\begin{equation}
	\nu_\infty=\nu_e \, e^{\alpha(r_e)-\alpha(r_\infty)} \left(E e^{-\alpha(r_e)}\mp \frac{J}{r_e}\right)^{-1}.
\end{equation}   
This frequency change includes that of the photon climbing out the gravitational potential (i.e. its gravitational redshift); to extract out the change in frequency due to the motion of the star, we simply compute the \emph{difference} in the observed frequency as the star moves towards  and away from us. Using equation (\ref{eq:dopcond1}) we find that to lowest order in small parameters this is 
\begin{equation}
	\frac{\Delta \nu}{\nu_e}\equiv \frac{\nu_{\infty+}-\nu_{\infty-}}{2\nu_e}\approx\frac{J}{r_e}, 
	\label{eq:rotationcurve}
\end{equation}
which an observer interprets as being due to the peculiar velocity $v$ of the star, 
\begin{equation} \label{eq:velodefi}
	v=\frac{\Delta\nu}{\nu_e}=\frac{J}{r_e}.
\end{equation}
Using equation (\ref{eq:doppler}) To first order in $J/r$ and dropping the subscript $e$, the rotation curve (\ref{eq:velodefi}) becomes 
\begin{equation} \label{eq:rotcurve}
	v^2(r)=\frac{G m(r)}{r}+4\pi G r^2 p(r).
\end{equation}
Hence, once we have a solution for the energy density $\rho$, and a relation between $\rho$ and the pressure $p$, we can simply insert them into the right hand side of the above equation to obtain the rotation curve. 

\section{Stability and Initial Value Formulation} \label{sec:stability}

In the previous sections we have studied several different static configurations of spherically symmetric non-canonical scalar fields. Presumably, these are the end-result of the gravitational collapse of k-field matter. If so, they should be stable, since otherwise they would not be the end stage of a dynamical process. The stability of scalar field configurations depends on the microscopic properties of the scalar, i.e. on its Lagrangian $L$. In the context of cosmological solutions (where the scalar field gradient is timelike), classical stability requires that the speed of sound of the k- field be positive \cite{GarrigaMukhanov},
\begin{equation}\label{eq:soundspeed}
	c_s^2=\frac{dp_c}{d\rho_c}=\frac{dL/dX}{dL/dX+2Xd^2L/dX^2}>0\quad\quad (X>0).
\end{equation}
Note that this is a necessary (but not sufficient) condition, since in principle we can still suffer from instabilities in the long wavelength regime\footnote{We thank Raul Abramo for bringing this point up.}.

On the other hand, at the quantum level, stability requires the Hamiltonian of field perturbations to be positive definite \cite{ghost}, which translates into
\begin{equation}\label{eq:QMstability}
	\frac{dL}{dX}>0 \quad \text{and} \quad \frac{dL}{dX}+2 X \frac{d^2L}{dX^2}>0
	\quad\quad (X>0).
\end{equation}
If the last conditions are not satisfied, quantum fluctuations around the background cosmological solution have negative energy, and the vacuum can decay into positive energy gravitons and negative energy scalar field perturbations \cite{CaHoTr}. Note that quantum stability implies classical stability.

In the first subsection of this appendix, we analyze both classical and quantum stability of static, spherically symmetric configurations of a k-field. Again, we find that quantum stability implies classical stability. Moreover, it turns out that the stability conditions we find are intimately related to the existence of a well-posed initial value formulation, which we study in the second subsection of the appendix.  As opposed to  studies of stability, our approach here is quite general;  we do not  make any assumption about the nature of the scalar field or the spacetime metric, and we do not rely on perturbation theory.

\subsection{Stability}

Consider a static spherically solution $X_0(r)$ of the equation of motion (\ref{eq:conservation}). We would like to find out whether this solution is quantum-mechanically stable, i.e. whether it can decay through quantum processes. Because any arbitrary spacetime is locally flat, we can assume that locally the metric is Minkowski.  Moreover, because we are studying static solutions the gradient field is spacelike,  locally we can also assume that the field gradient points in a given spatial direction, say, the $z$ direction, $\varphi=\sqrt{-2X_0}z$. Let us perturb this local expression,  
\begin{equation}
	\varphi=\sqrt{-2X_0}\, z+ \delta\varphi,
\end{equation}
and study the resulting  Lagrangian. Using the definition (\ref{eq:squaredgradient}) we find that the perturbed gradient is
\begin{equation}
	X=X_0-\sqrt{-2X_0}\partial_z \delta\varphi
	+\frac{1}{2}\partial_\mu \delta\varphi \partial^\mu\delta\varphi,
\end{equation}
where we are using the Minkowski metric. Hence, expanding the Lagrangian $L(X)$ to quadratic order in
$\delta\varphi$ we find
\begin{equation}\label{eq:pertLagrangian}
	\delta L=\frac{1}{2}\left[L_{,X}(X_0) \delta\varphi_{,t}^2
	-L_{,X}(X_0)(\delta\varphi_{,x}^2+\delta\varphi_{,y}^2)
	-(2X_0 L_{,XX}(X_0)+L_{,X}(X_0)) \delta\varphi_{,z}^2\right],
\end{equation}
where a subscript $,i$ denotes a partial derivative with respect to the coordinate variable $i$. The Hamiltonian of the scalar perturbations is positive definite only if the signs of the quadratic derivative terms in the Lagrangian are positive for time derivatives and negative for spatial derivatives. Thus, the condition for quantum mechanical stability is the same as in a cosmological background, equations (\ref{eq:QMstability}).

Although equation (\ref{eq:pertLagrangian}) already suffices to determine the propagation speed of classical perturbations, it is more informative to consider an approach that maintains the symmetries of the background.  Hence, we expand a scalar field perturbation $\delta\varphi$ in spherical harmonics $Y_{lm}$,
\begin{equation}
	\delta\varphi=\sum_{l,m} \delta\varphi_{lm}(t,r) Y_{lm}(\theta,\phi),
\end{equation}
and instead of expanding the Lagrangian to quadratic order, we linearize the equation of motion. Because of linearity,  it is sufficient to consider one spherical mode at a time. Moreover, by rotational symmetry, we can always assume $m=0$. Since matter is coupled to geometry through Einsteins equations, linear perturbations in the scalar field should source metric perturbations. However, we do not expect the speed at which small perturbations propagate to depend on the strength of such coupling, Newton's constant $G$. So we simplify our problem by considering the limit $G\to 0$.\footnote{Strictly speaking only limits of dimensionless quantities are physically meaningful. By $G\to 0$ we mean the limit where  every dimensionless quantity proportional to $G$ goes to zero.} In this limit matter perturbations do not source any metric perturbations, and one can simply study the propagation of the scalar field in a given, fixed, gravitational background. In this way, substituting $\varphi=\varphi_0(r)+\delta\varphi$  into the equation of motion (\ref{eq:conservation}) and keeping terms linear in $\delta\varphi$ we arrive at 
\begin{equation}\label{eq:linearized}
 	\frac{1}{e^{2\alpha}}\frac{d^2\delta\varphi_l}{dt^2}
	-\frac{2XL_{,XX}+L_{,X}}{L_{,X}}\frac{d^2\delta\varphi_l}{e^{2\beta}dr^2} 
	-\frac{\frac{d}{dr}\left[r^2 e^{\alpha-\beta}(2XL_{,XX}+L_{,X})\right]}
	{e^\beta[r^2 e^{\alpha-\beta} L_{,X}]}\frac{d\delta\varphi_l}{e^\beta dr}
	+\frac{l(l+1)}{r^2}\delta\varphi_l =0 
\end{equation} 
Note that a static observer at radius $r$ measures proper time $d\tau=e^\alpha dt$ and proper distance $ds=e^\beta dr$. By inspection one can see then that the speed of propagation of radial perturbations is
\begin{equation}\label{eq:cssq}
	c_s^2=\frac{2XL_{,XX}+L_{,X}}{L_{,X}}\quad\quad (X<0).
\end{equation}
This is the \emph{inverse} of squared the speed of sound in a cosmological setting (where $X$ is timelike), equation (\ref{eq:soundspeed}). Indeed, the definition of the squared speed of sound $c_s^2=dp/d\rho$ agrees with equation (\ref{eq:cssq}) if the field is spacelike and $p$ and $\rho$ are given by equations (\ref{eq:p}), and it is the inverse  of equation (\ref{eq:cssq}) if the field is timelike and $p_c$ and $\rho_c$ are given by equation (\ref{eq:rho}). The linearized equation of motion also shows that for non-radial perurbations the speed of sound is $c_s^2=1$, which, up to a sign, also agrees with the expectation $c_s^2=dp/d\rho$. Note that the sound speed is given by equation (\ref{eq:cssq}) even if the Lagrangian explicitly depends on $\varphi$, i.e. $L=L(X,\varphi)$.

\subsection{Initial Value Formulation}
In this part of the Appendix we determine for what functions $L$ the scalar field equation possesses a well-posed initial value formulation. After all, we are mainly dealing with solutions of a classical field theory, and requiring the equations to have a well-posed initial value formulation amounts to self-consistency of our approach. As mentioned above,  we will not  make any assumption about the nature of the scalar field or the metric, and we will not rely on perturbation theory.

We begin by rewriting the k-field equation of motion (\ref{eq:conservation}) as 
\begin{equation}\label{eq:hyperbolic}
	h^{\mu\nu}\nabla_\mu \nabla_\nu \varphi=0,
\end{equation}
where the ``induced'' metric is
\begin{equation}
	h^{\mu\nu}=\frac{dL}{dX} g^{\mu\nu}+\frac{d^2L}{dX^2}\nabla^\mu\varphi\nabla^\nu 	\varphi.
\end{equation}
For simplicity we assume that $L$ only depends on $X$, though our results also apply if the Lagrangian
 depends on $\varphi$ itself. It can be proven (see Theorem 10.1.3. of \cite{Wald}) that if the metric $h^{\mu\nu}$ is Lorentzian, the field equation  admits a well-posed initial value-formulation (at least locally), so all we have to show is that $h^{\mu\nu}$ has a negative determinant. 

Let $\mathbf{h}$, $\mathbf{g}$ and $\mathbf{n}$ denote the matrices with elements  $h^{\mu\nu}$, $g^{\mu\nu}$ and $\nabla^\mu\varphi\nabla^\nu\varphi$ respectively. We then have
\begin{equation}
	\det \mathbf{h}=\det\left(\frac{dL}{dX}\mathbf{g}+\frac{d^2L}{dX^2}\mathbf{n}\right)=
	\left(\frac{dL}{dX}\right)^4 \cdot
	\det \mathbf{g}\cdot \det \left(\mathbf{1}+\frac{d^2 L/dX^2}{dL/dX} \mathbf{g}^{-1} \mathbf{n}\right),
\end{equation}
where $\mathbf{g}^{-1}$ denotes the inverse of $\mathbf{g}$, and we have assumed $dL/dX\neq 0$. 
Using the identity $\det e^{\mathbf{A}}=e^{\text{tr} \mathbf{A}}$ we find
\begin{equation}\label{eq:identity}
	\det \left(\mathbf{1}+\frac{d^2 L/dX^2}{dL/dX} \mathbf{g}^{-1} \mathbf{n}\right)=
	\exp\left[\text{tr} \ln\left(\mathbf{1}+ \frac{d^2 L/dX^2}{dL/dX} \mathbf{g}^{-1}\mathbf{n}\right)\right],
\end{equation}
and expanding the logarithm in a Taylor series around the identity we obtain 
\begin{equation}\label{eq:Tln}
	\text{tr}\, \ln\left(\mathbf{1}+ \frac{d^2 L/dX^2}{dL/dX} \mathbf{g}^{-1} \mathbf{n}\right)=
	\sum_k \frac{(-1)^{k+1}}{k} \left(\frac{d^2 L/dX^2}{dL/dX}\right)^k
	\text{tr} \left(\mathbf{g}^{-1}\mathbf{n}\right)^k=\ln \left(1+2X\frac{d^2 L/dX^2}{dL/dX}\right),
\end{equation}
where we have used  $\text{tr} \left(\mathbf{g}^{-1}\mathbf{n}\right)^k=(2X)^k$. Finally, substituting equation (\ref{eq:Tln}) back into equation (\ref{eq:identity}) we arrive at our desired result 
\begin{equation}
	\det \mathbf{h}=\left(\frac{dL}{dX}\right)^4\cdot 
	\det \mathbf{g}\cdot \frac{dL/dX+2X d^2L/dX^2}{dL/dX}.
\end{equation}
By assumption, the metric of spacetime is Lorentzian, so that $\det \mathbf{g}<0$. Therefore,
$\mathbf{h}$ is Lorentzian if and only if 
\begin{equation}\label{eq:Lorentzian}
	\Delta^2\equiv \frac{dL/dX}{2X d^2L/dX^2+dL/dX}>0. 
\end{equation}

So far, we have found that the metric $h^{\mu\nu}$ is Lorentzian if condition (\ref{eq:Lorentzian}) is satisfied. Because the assumptions of the theorem mentioned above apply, the k-field equation of motion is guaranteed to have a well-posed initial value formulation, at least locally. An ingredient of such a well-posed formulation is the ``causal" propagation of the scalar field.  Note however that the metric that enters equation (\ref{eq:hyperbolic}) is not the space-time metric $g^{\mu\nu}$, but the derived metric $h^{\mu\nu}$. Therefore, causality in the propagation of the k-field means that signals propagate inside the light-cone of the metric $h$, and not $g$ \cite{Gibbons}.  This observation sheds some light on a property of non-canonical fields that was noted early on \cite{GaMuOlVi}. If the Lagrangian $L$ is an arbitrary function of $X$ it is possible to obtain superluminal sound-speeds, $c_s^2>1$. More generally, k-fields are natural implementations of bimetric theories, and they  might offer  a consistent theoretical framework to study and implement some of the ideas about causality as suggested in \cite{VSL}.

To further illustrate the causal structure tied to the propagation of a k-field, let us consider the inverse of the matrix $\mathbf{h}$, whose components we denote by $\tilde{h}_{\mu\nu}$,
\begin{equation}\label{eq:h}
	\tilde{h}_{\mu\nu}=\frac{1}{dL/dX}\left(g_{\mu\nu}-\frac{d^2L/dX^2}{dL/dX+2X d^2L/dX^2}\nabla_\mu\varphi\nabla_\nu\varphi\right).
\end{equation}  
Note that the inverse of $h^{\mu\nu}$ is not given by $h_{\mu\nu}\equiv g_{\mu\rho}\, g_{\nu\sigma}h^{\rho\sigma}$.   The metric $\tilde{h}_{\mu\nu}$ determines the causal structure of the solutions to equation (\ref{eq:hyperbolic}), and the theorem mentioned above guarantees that no signal can travel outside the light cone defined by $\tilde{h}_{\mu\nu}$. The question we are interested in is the following. Suppose that a vector field $v^\mu$ is null with respect to $h$ ($\tilde{h}_{\mu\nu}v^\mu v^\nu$=0), is $v^{\mu}$ timelike, spacelike or also null with respect to $g$? If $v^\mu$ is null with respect  to $\tilde{h}_{\mu\nu}$, it follows from equations (\ref{eq:Lorentzian}) and (\ref{eq:h})  that
\begin{equation}
	g_{\mu\nu}v^\mu v^\nu=(1-\Delta^{2})\frac{(\nabla_\mu\varphi \,v^\mu)^2}{2X}.
\end{equation}
Hence, there are two cases, as in Fig. \ref{figcone}: 
\begin{itemize}

\item{Timelike gradient ($X>0$)}

In this case the vector $v^\mu$ is  timelike for $\Delta^2<1$ and spacelike for $\Delta^{2}>1$. For $\Delta^2<1$ the light cones of the metric $h$ are contained in the light cones of $g$ and for $\Delta^2>1$, the light cones of $g$ are contained in the light cones of $h$, as one expects. This is consistent with  $\Delta^{2}$ being  the squared  sound speed, 
\begin{equation}
	c_{s}^{2}=\Delta^{2}.
\end{equation}

\begin{figure}[ptbh]
	\myfigure{4in}{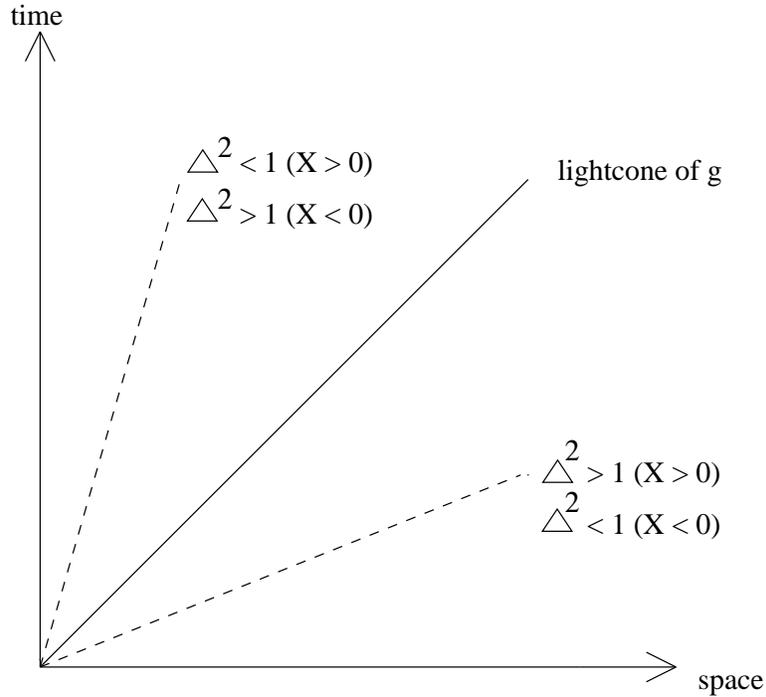}
	\caption{The lightcones of $g$ and $h$ for the given conditions on $X$ and $\Delta$. The scalar field propagates inside the lightcones of $h$.} \label{figcone}
\end{figure}

\item{Spacelike gradient ($X<0$)}

In this case, the opposite happens. For $\Delta^{2}>1$ the light cones of the metric $h$ are contained in the light cones of $g$ and for $\Delta^2<1$, the light cones of $g$ are contained in the light cones of $h$. This is also consistent with our observation that in a static, spherically symmetric spacetime radial perturbations propagate with squared speed,
\begin{equation}
	c_{s}^{2}=\Delta^{{-2}}.
\end{equation}
\end{itemize}

\section{Non-singular solutions: de Sitter spacetime}\label{sec:nonsingular}

Our solutions in the weak field limit of the main text are all singular at the origin ${r=0}$. Therefore, our haloes appear to  have density cusps at their centers. Of course, near the origin the weak field conditions, equations (\ref{eq:Weakfield}), are violated, so we do not expect our solutions to be valid in this regime. It is interesting to ask whether these singularities are an artifact of the weak field approximation, or whether they still persist in solutions of the full non-linear equations. Here, by singular we mean that in the coordinate system defined by equation (\ref{eq:metric}) quantities like the mass of the halo, the pressure, the energy density and the metric coefficients diverge. In this section, we show that the only one non-singular solution of the Einstein equations coupled to a k-field whose Lagrangian only depends on $X$ is de Sitter spacetime. We note in passing that there are other non-singular static, spherically symmetric solutions to these equations, but they typically describe wormholes \cite{wormholes} instead of haloes, and hence, cannot be described in the coordinates we are using.

Let us begin with the full Tolman-Oppenheimer-Volkoff equations (\ref{eq:TOV}).  We are going to look for regular solutions, so we can assume that $m$, $\rho$ and $p$ admit Taylor expansions around $r=0$,
 \begin{eqnarray}
 	m&=&m_1r+m_2r^2+\cdots \label{eq:Tmass} \\
 	\rho&=&\rho_0+\rho_1r+\cdots \label{eq:Tp}\\
 	p&=&p_0+p_1r+\cdots, \label{eq:Ttau}
 \end{eqnarray}
where we have used that in order for the metric to be non-singular, $m_0=0$. Because both $\rho$ and $p$ depend on $X$, the squared gradient can be similarly expanded
\begin{equation}
	X=X_0+X_1r+\cdots.
\end{equation}
Now, we insert these expansions back into the TOV equations (\ref{eq:TOV}) and do a term by term comparison. Substituting equation (\ref{eq:Tmass}) into equation (\ref{eq:dm}) and comparing coefficients we find 
 \begin{equation}\label{eq:masscoeff}
 	m_1=m_2=0.
 \end{equation}
Also, inserting equation (\ref{eq:Tmass}) into equation (\ref{eq:dalpha}), one sees than $d\alpha/dr$ is at most linear in $r$. Meanwhile, we substitute the expansions (\ref{eq:Tp}) and (\ref{eq:Ttau}) into the third TOV equation (\ref{eq:motion}) and compare coefficients, finding to $\mathcal{O}(r^{-1})$ that $\rho_0+p_0=0$.  Therefore, using equation (\ref{eq:p}), we find that either
 \begin{equation}\label{eq:zero}
 	X_0=0 \quad \text{or} \quad \frac{dL}{dX}\Big|_{X_0}=0.
 \end{equation}
Using the same TOV equation, and armed with our knowledge that $d\alpha/dr=\mathcal{O}(r)$, we obtain the relation $2\rho_1+3p_1=0$.
Now there are two possibilities; if $X_0\neq 0$, then by equation (\ref{eq:zero}), $\rho_1=-L_{,X}(X_0)X_1=0$, which implies $\rho_1=0$.  On the other hand, if $X_0=0$ we find that $2\rho_1+3p_1=-5L_{,X}(X_0)X_1=5\rho_1$ has to vanish, which in turn implies the vanishing of $p_1$.

We can use these results to set initial conditions at a sufficiently small non-vanishing $r=\Delta r$. Because $\rho_0+p_0=0$, and because $\rho_1=p_1=0$, we find that $\rho(\Delta r)+p(\Delta r)=0.$ With this initial conditions we can integrate equation (\ref{eq:motion}) from $r=\Delta r$ onwards immediately, to find that $p$ and $\rho$ are both constants. This is the only non-singular solution. What is the nature of the non-singular solution we have found? Since $\rho$ is constant and positive for all $r\geq 0$, the spacetime is simply de Sitter. To verify this is indeed true, we integrate equations (\ref{eq:dm}) and (\ref{eq:dalpha}) in the following. 

Our result $p=-\rho=\mathrm{constant}$ already solves (\ref{eq:motion}) for all $r$ in one fell swoop. We can then integrate (\ref{eq:dm}) to obtain 
\begin{equation}
	m=\frac{4\pi}{3} r^3\, \rho_0,	 
\end{equation}
which once substituted into equation (\ref{eq:dalpha}) yields 
 \begin{equation}
 	\alpha=\frac{1}{2}\log\left(1-\frac{8\pi G}{3}r^2 \rho_0\right),
\end{equation}
where we have assumed $\alpha(0)=0$. Putting everything back together, the metric is then
\begin{equation}
	ds^2=\left(1-\frac{8\pi G}{3}r^2 \rho_0\right)dt^2-\left(1-\frac{8\pi G}{3}r^2 \rho_0\right)^{-1}dr^2 -r^2 d\Omega^2
\end{equation}
which is simply de Sitter space in static coordinates. 

It has been known for a while that de Sitter solutions arise for timelike field gradients in k-field theories. These solutions exist if the Lagrangian has an extremum at a positive value of $X$, $dL/dX=0$ at $X_0>0$. However, for such Lagrangians the term proportional to $X$ typically has the opposite sign \cite{k-inflation}. The constraints that the existence of de Sitter solutions for spacelike values of the squared gradient imposes on the Lagrangian are drastically different.    For example, the conventionally-looking  Lagrangian $L=X+X^2$ does not have a point where  $dL/dX=0$ at $X>0$ (timelike), but does have a point where $dL/dX=0$ at $X<0$ (spacelike).  Hence, de Sitter solutions at spacelike $X$ might offer  an alternative interesting possibility to implement inflation and cosmic accelerations with ``conventional" Lagrangians.  In this context, it is instructive to look at our spacelike solution in the coordinate system where de Sitter has the form of a flat FRW universe,
\begin{equation}
	ds^2=d\tilde{t}^2-\exp(2H\tilde{t})[d\tilde{r}^2+\tilde{r}^2 (d\theta^2+\sin^2\theta d\phi^2].
\end{equation}
In this coordinate system the scalar field is given by
\begin{equation}
	\varphi=\sqrt{-2X_0} H^{-1} \arcsin\left(H\tilde{r}e^{H\tilde{t}}\right),
\end{equation}
and it can be easily shown that such a $\varphi$ leads to a constant $X=X_0<0$. The scalar field is inhomogeneous, and becomes ill-defined at the horizon, $\tilde{r}e^{H\tilde{t}}=H^{-1}$.

\end{document}